\documentclass[aps,onecolumn,nobalancelastpage,amsmath,amssymb,nofootinbib,groupedaddress]{revtex4}

\usepackage{blindtext}
\usepackage{natbib}
\usepackage[T1]{fontenc}
\usepackage[english]{babel}
\usepackage{amsmath}
\usepackage{amsfonts}
\usepackage{amssymb}
\usepackage[colorlinks=true,linkcolor=blue,urlcolor=blue,citecolor=blue]{hyperref}
\usepackage{lineno}
\usepackage{braket}
\usepackage{wrapfig}
\usepackage{todonotes}

\begin{document}
\title{Time dependent coupled harmonic oscillators}
%%%%%%%%%%%
\author{Alejandro R. Urz\'ua}
\affiliation{Instituto Nacional de Astrof\'isica, \'Optica y Electr\'onica, Calle Luis Enrique Erro No. 1, Santa Mar\'ia Tonantzintla, Puebla, 72840, Mexico}
\author{Ir\'an Ramos-Prieto}
\affiliation{Instituto de Ciencias F\'isicas, Universidad Nacional Aut\'onoma de M\'exico, Apartado Postal 48-3, 62251 Cuernavaca, Morelos, Mexico}
\email[Corresponding author: ]{hmmc@inaoep.mx}
\author{Manuel Fern\'andez Guasti}
\affiliation{Departamento de F\'isica, CBI,
Universidad Aut\'onoma Metropolitana  Iztapalapa,
Apartado Postal 55-534, M\'exico D.F. 09340, Mexico}
\author{ H\'ector M. Moya-Cessa}
\affiliation{Instituto Nacional de Astrof\'isica, \'Optica y Electr\'onica, Calle Luis Enrique Erro No. 1, Santa Mar\'ia Tonantzintla, Puebla, 72840, Mexico}

\begin{abstract}
We show that, by using the quantum orthogonal functions invariant, we are able to solve a coupled of time dependent harmonic oscillators where all the time dependent frequencies are arbitrary. We do so, by transforming the time dependent Hamiltonian of the interaction by a set of unitary operators. In passing, we show that $N$ time dependent and coupled oscillators have a generalized orthogonal functions invariant from which we can write a Ermakov-Lewis invariant.
\end{abstract}

\date{\today}
\maketitle

\section{Introduction}

The existence of invariants in mechanical systems for time dependent Hamiltonian has attracted considerable interest over the years \cite{Bouquet and
Lewis}. Such constants of motion are of central importance in the study of dynamical systems. A  variety of methods to obtain invariants of systems with one degree of freedom have been developed  \cite{Ray and
Reid}. In particular, the time dependent harmonic oscillator (TDHO) has received much attention bacause of its applications in several areas of physics
\cite{Colegrave}. Among the many procedures developed to obtain invariants, a derivation for the classical TDHO has been presented, that leads directly to the orthogonal functions invariant or to the Lewis invariant \cite{fer1}. The study of exact invariants has led to the
nonlinear superposition principle as well as the obtention of general
solutions provided that a particular solution is known.

The extension of the theory of invariants to the quantum realm has evolved in,
at least, two directions. On the one hand, the one dimensional time
independent Schr\"{o}dinger equation is formally equivalent to the TDHO
equation. The translation between equations requires the exchange of temporal
and spatial variables as well as a constant shift of the potential $V\left(
x\right)  $ with the appropriate scaling for the initially time dependent
parameter $\Omega^{2}(t)\rightarrow{2m}\left(  \mathcal{E}%
-V\left(  x\right)  \right)  $. The results obtained in the classical
invariant theory are thus applicable for spatially arbitrary time independent
potentials in stationary one dimensional quantum theory. Using these technique it has been possible to define coherent states associated to the TDHO \cite{Moya2003} and amplitude-phase invariants \cite{Guasti2003b}.  On the other hand,
quantum mechanical expressions of the classical invariant operators have been
used in order to obtain exact solutions to the time dependent Schr\"{o}dinger
equation. To this end, the classical Hamiltonian is translated into a quantum
Hamiltonian by considering the canonical coordinate and momentum as time
independent operators obeying the commutation relationship $\left[  \hat
{q},\hat{p}\right]  =i$ (we will set $\hbar=1$ throughout the manuscript). The quantum treatment becomes then a 1+1
dimensional problem where the wave function depends on a spatial as well as
the temporal variable. A potential with an arbitrary time dependence is
identified with the coordinate operator of the Hamiltonian. Exact invariants
have been derived to tackle a limited class of admissible potentials
\cite{Lewis and Leach}. The most relevant cases are the linear potential \cite{Guedes} and the quadratic spatial
dependence that leads to the quantum mechanical time dependent harmonic oscillator (QM-TDHO).

On the other hand, the simple extension to two coupled time dependent harmonic oscillators has been considered and solution to it have been presented for a very limited case of time dependent functions \cite{Macedo2012}. Ermakov-Lewis invariant have been also proposed for systems of couple harmonic oscillators \cite{Thylwe1998}.

The QM-TDHO has been solved under various scenarios such as  time dependent mass \cite{Moya2007,Ramos2018} and damping \cite{Yeon}. Several techniques have been used to solve the
corresponding time dependent Schr\"{o}dinger equation such as the time-space re-scaling or transformation method and the time dependent invariant method
\cite{Ray}. The constant of motion that has been invoked in the latter procedure is the well known Lewis invariant \cite{Lewis}. 

The main purpose of the present contribution is to show a method to solve the Schr\"odinger equation for a pair of coupled time dependent harmonic oscillators when all the time dependent functions involved are arbitrary, i.e., they are not related to each other. In passing, we write the Ermakov-Lewis invariant for  $N$ coupled time dependent harmonic oscillators.

By a series of unitary transformations, some of them, time dependent, we manage to take the Hamiltonian for the two coupled harmonic oscillators to an integrable form.
\section{Ermakov-Lewis invariant for $N$ coupled time dependent harmonic oscillators}
Consider the system of differential equations for $N$ time dependent coupled classical oscillators
\begin{align}
\begin{split}
\ddot{u}_1+\Omega_1^2(t)u_1&=-\eta_{12}(t)u_2\\
\ddot{u}_2+\Omega_2^2(t)u_2&=-\eta_{12}(t)u_1-\eta_{23}(t)u_3\\
\ddot{u}_3+\Omega_3^2(t)u_3&=-\eta_{23}(t)u_2-\eta_{34}(t)u_4 \\ 
\vdots& \\ 
\ddot{u}_N+\Omega_N^2(t)u_N&=-\eta_{N-1N}(t)u_{N-1}, \label{sistema}
\end{split}
\end{align}
with the associated quantum Hamiltonian 
\begin{align}
\begin{split}
\hat{H}_N(t)&=\frac{1}{2}\left({\hat{p}_1^2}+{\Omega_1^2(t)\hat{x}_1^2}\right)
+\frac{1}{2}\left({\hat{p}_2^2}+{\Omega_2^2(t)\hat{x}_2^2}\right)+\dots+ \frac{1}{2}\left({\hat{p}_N^2}+{\Omega_N^2(t)\hat{x}_N^2}\right)\\&+ \eta_{12}(t)\hat{x}_1\hat{x}_2+ \eta_{23}(t)\hat{x}_2\hat{x}_3+\dots +\eta_{N-1N}(t)\hat{x}_{N-1}\hat{x}_N. \label{Mosc}
\end{split}
\end{align}
A single time dependent harmonic oscillator has quantum orthogonal functions invariant \cite{JPA},
\begin{equation}
\hat{G}_1=u_1(t)\hat{p}_1-\dot{u}_1(t)\hat{x}_1, \label{inva}%
\end{equation}
where $u_1(t)$ is the solution of the equation $\ddot{u}_1+\Omega_1^2(t)u_1=0$. This invariant may be generalized to $N$ coupled time dependent harmonic oscillators,
\begin{equation}
\hat{G}_N=u_1(t)\hat{p}_1-\dot{u}_1(t)\hat{x}_1+u_2(t)\hat{p}_2-\dot{u}_2(t)\hat{x}_2+\dots+
u_N(t)\hat{p}_N-\dot{u}_N(t)\hat{x}_N,  \label{Ninvariant}
\end{equation}
where the $u$'s satisfy \eqref{sistema}, such that 
\begin{eqnarray}
\frac{\partial \hat{G}_N}{\partial t}=\dot{u}_1(t)\hat{p}_1-\ddot{u}_1(t)\hat{x}_1+\dot{u}_2(t)\hat{p}_2-\ddot{u}_2(t)\hat{x}_2+\dots+
\dot{u}_N(t)\hat{p}_N-\ddot{u}_N(t)\hat{x}_N.
\end{eqnarray}
On the other hand, the commutator between  $\hat{G}_N$ and $\hat{H}_N$, is
\begin{align}
\begin{split}
i[\hat{G}_N,\hat{H}_N]&=\Omega_1^2(t)u_1\hat{x}_1+\dot{u}_1(t)\hat{p}_1+\eta_{12}u_1\hat{x}_2\\&+
\Omega_2^2(t)u_2\hat{x}_2+\dot{u}_2(t)\hat{p}_2+\eta_{12}u_2\hat{x}_1+\eta_{23}u_2\hat{x}_3\\ &\vdots
\\ &+
\Omega_N^2(t)u_N\hat{x}_N+\dot{u}_N(t)\hat{p}_N+\eta_{N-1N}u_N\hat{x}_{N-1},
\end{split}
\end{align}
by subtracting the above equations we obtain
\begin{align}
\begin{split}
\frac{d\hat{G}_N}{dt}=\frac{\partial \hat{G}_N}{\partial t}-i[\hat{G}_N,\hat{H}_N]&=-[\Omega_1^2(t)u_1+\ddot{u}_1]\hat{x}_1-\eta_{12}u_1\hat{x}_2\\&-
[\Omega_2^2(t)u_2+\ddot{u}_2]\hat{x}_2-\eta_{12}u_2\hat{x}_1-\eta_{23}u_2\hat{x}_3\\&\vdots\\ &-
[\Omega_N^2(t)u_N+\ddot{u}_N]\hat{x}_N-\eta_{N-1N}u_N\hat{x}_{N-1}. 
\end{split}
\end{align}
Rearranging the above expression
\begin{align}
\begin{split}
\frac{d\hat{G}_N}{dt}=\frac{\partial \hat{G}_N}{\partial t}-i[\hat{G}_N,\hat{H}_N]&=-[\Omega_1^2(t)u_1+\ddot{u}_1+\eta_{12}(t)u_2]\hat{x}_1\\&-
[\Omega_2^2(t)u_2+\ddot{u}_2+\eta_{12}(t)u_1+\eta_{23}(t)u_3]\hat{x}_2\\&\vdots \\&-
[\Omega_N^2(t)u_N+\ddot{u}_N+\eta_{N-1N}(t)u_{N-1}]\hat{x}_N 
\end{split}
\end{align}
that from (\ref{sistema}) gives  zero showing that $\hat{G}_N$ is indeed an invariant.

If we write, for the single harmonic oscillator $u_1=\rho_1\exp(-i\int \frac{dt}{\rho_1^2})$, where $\rho_1$ obeys the Ermakov equation \cite{JPA}
\begin{equation}
\ddot{\rho}_1+\Omega_1^2(t)\rho_1=\frac{1}{\rho_1^3}.
\end{equation}
The so-called Ermakov-Lewis invariant may be obtained from $\hat{G}_1$ as
\begin{equation}
\hat{I}_1=\hat{G}_1\hat{G}_1^{\dagger}=\frac{1}{2}\left(\frac{\hat{x}_1^2}{\rho_1^2}+(\rho_1\hat{p}_1-\dot{\rho_1}\hat{x}_1)^2\right),
\end{equation}
such that we may write the Ermakov-Lewis invariant for the $N$ coupled time dependent harmonic oscillators as
\begin{equation}
\hat{I}_N=\hat{G}_N\hat{G}_N^{\dagger}.
\end{equation}
\subsection{The classical invariant}
By doing $\hat{p}\rightarrow \dot{v}$ and $\hat{x}\rightarrow {v}$ in (\ref{Ninvariant}) we find the classical invariant
\begin{equation}
{G}_N=u_1(t)\dot{v}_1(t)-\dot{u}_1(t){v}_1(t)+u_2(t)\dot{v}_2(t)-\dot{u}_2(t){v}_2(t)+\dots+
u_N(t)\dot{v}_N(t)-\dot{u}_N(t){v}_N(t),
\end{equation}
where the $u$'s and $v$'s are linearly independent solutions of (\ref{sistema}).

\section{Two-coupled time dependent harmonic oscillators}
We consider the time dependent Hamiltonian for the interacting oscillators as
\begin{equation}
\hat{H}(t)=\frac{1}{2}\left[\hat{p}_x^2+\hat{p}_y^2+\Omega_x^2(t)\hat{x}^2+\Omega_y^2(t)\hat{y}^2\right]+\eta(t)\hat{x}\hat{y},
\end{equation}

The classical equations of motion for the above Hamiltonian are 
\begin{equation}\label{difeq}
\ddot{u}_x+\Omega_x^2(t)u_x=-\eta(t)u_y, \qquad   \ddot{u}_y+\Omega_y^2(t)u_y=-\eta(t)u_x,
\end{equation}

where the quantum invariants of each coupled oscillator are \cite{MFGPL}
\begin{equation}
\hat{G}_{x}=\left(u_{x}\hat{p}_{x}-\hat{x}\dot{u}_{x}\right)-\int\eta(t)\left(u_{x}\hat{y}-\hat{x}u_{y}\right)dt,\label{eq:Q00 ux quant inv}
\end{equation}
and
\begin{equation}
\hat{G}_{y}=\left(u_{y}\hat{p}_{y}-\hat{y}\dot{u}_{y}\right)-\int\eta(t)\left(u_{y}\hat{x}-\hat{y}u_{x}\right)dt,\label{eq:Q00 uy quant inv}
\end{equation}
since the total invariant must comply with
\begin{equation}
    \frac{\partial}{\partial t}(\hat{G}_x+\hat{G}_y)-i[\hat{G}_x+\hat{G}_y,\hat{H}]=0.
\end{equation}

We now consider the transformation \cite{JPA}
\begin{equation}
\hat{T}_{u}=e^{i\frac{\ln u_x(t)}{2 }(\hat{x}\hat
{p}_x+\hat{p}\hat{x})}e^{-i\frac{\dot{u}_x(t)}{2  u_x(t)}\hat{x}^{2}}e^{i\frac{\ln u_y(t)}{2 }(\hat{y}\hat{p}_y+\hat{p}_y\hat{y})}e^{-i\frac{\dot{u}_y(t)}{2  u_y(t)}\hat{y}^{2}}
\end{equation}
that produces
\begin{align}
\begin{split}
\hat{T}_{u}\hat{x}\hat{T}_{u}^{\dagger}&=u_x\hat{x} \\
\hat{T}_{u}\hat{y}\hat{T}_{u}^{\dagger}&=u_y\hat{y} \\
\hat{T}_{u}\hat{p}_x\hat{T}_{u}^{\dagger}&=\frac{\hat{p}_x}{u_x}+{\dot{u}_x}\hat{x}, \\
\hat{T}_{u}\hat{p}_y\hat{T}_{u}^{\dagger}&=\frac{\hat{p}_y}{u_y}+{\dot{u}_y}\hat{y},
\end{split}
\end{align}
where $u$ is the solution to TDHO Eq. (\ref{difeq}).

If we transform the wave function with the transformation above, i.e., 
\begin{equation}
|\phi_{u}(t)\rangle=\hat{T}_{u}|\psi(t)\rangle
\end{equation}
the Schr\"{o}dinger equation 
\begin{equation}
i \frac{\partial|\psi(t)\rangle}{\partial t}=\hat{H}(t)|\psi
(t)\rangle\label{scro}%
\end{equation}
has to be rewritten. In order to do it,
by substitution in the
above equation (\ref{scro}) leads to:
\begin{equation}
i\left(  \hat{T}_{u}^{\dagger}\frac{\partial|\phi_{u}(t)\rangle}{\partial
t}+\frac{\partial\hat{T}_{u}^{\dagger}}{\partial t}|\phi_{u}(t)\rangle\right)
=\hat{H}(t)\hat{T}_{u}^{\dagger}|\phi_{u}(t)\rangle.\label{tra}%
\end{equation}
By noting that
\begin{eqnarray}
\frac{\partial\hat{T}_{u}^{\dagger}}{\partial t}=\frac{i}{2}\hat{T}
_{u}^{\dagger}\left[  \left(  u_x\ddot{u}_x-\dot{u}_x^{2}\right)  \hat{x}^{2}
-\frac{\dot{u}_x}{u_x}(\hat{p}_x\hat{x}+\hat{x}\hat{p}_x)+\left(  u_y\ddot{u}_y-\dot{u}_y^{2}\right)  \hat{y}^{2}
-\frac{\dot{u}_y}{u_y}(\hat{p}_y\hat{y}+\hat{y}\hat{p}_y)\right]  ,
\end{eqnarray}
or 
\begin{eqnarray}
i\frac{\partial|\phi_{u}(t)\rangle}{\partial t}=\frac{1}{2}\left[
\frac{\hat{p}_x^{2}}{u_x^{2}}+\left(  \Omega_x^{2}u_x+\ddot{u}_x\right)  u_x\hat{x}%
^{2}+\frac{\hat{p}_y^{2}}{u_y^{2}}+\left(  \Omega_y^{2}u_y+\ddot{u}_y\right)  u_y\hat{y}%
^{2}+ \eta(t)u_xu_y\hat{x}\hat{y}\right]  |\phi_{u}(t)\rangle,
\end{eqnarray}
from (\ref{difeq}) we may rewrite the Schr\"odinger equation as
\begin{eqnarray}
i\frac{\partial|\phi_{u}(t)\rangle}{\partial t}=\frac{1}{2}\left[
\frac{\hat{p}_x^{2}}{u_x^{2}}+\frac{\hat{p}_y^{2}}{u_y^{2}} -\eta(t)u_x  u_y\left(\hat{x}-\hat{y}\right)^2\right]  |\phi_{u}(t)\rangle.
\end{eqnarray}
We now perform a second transformation,   $|\phi_{\theta}\rangle=\hat{R}_{\theta}|\phi_u\rangle$, with $\hat{R}_{\theta}=\exp[i\theta(\hat{x}\hat{p}_y-\hat{y}\hat{p}_x)]$
\begin{equation}
i\frac{\partial |\phi_{\theta}\rangle}{\partial t}= \hat{H}_{\theta}(t)|\phi_{\theta}\rangle
\end{equation}
such that the different operators are transformed according to
\begin{align}
\begin{split}
\hat{R}_{\theta}\hat{x}\hat{R}_{\theta}^{\dagger}&=\hat{x}\cos\theta-\hat{y}\sin\theta, \\
\hat{R}_{\theta}\hat{y}\hat{R}_{\theta}^{\dagger}&=\hat{y}\cos\theta+\hat{x}\sin\theta, \\
\hat{R}_{\theta}\hat{p}_x\hat{R}_{\theta}^{\dagger}&=\hat{p}_x\cos\theta-\hat{p}_y\sin\theta, \\
\hat{R}_{\theta}\hat{p}_y\hat{R}_{\theta}^{\dagger}&=\hat{p}_y\cos\theta+\hat{p}_x\sin\theta.
\end{split}
\end{align}
By settin $\theta=\pi/4$ to arrive to the integrable equation
\begin{eqnarray} \label{separada}
i\frac{\partial|\phi_{\theta}(t)\rangle}{\partial t}=\left[\frac{\hat{p}_x^{2}+\hat{p}_y^{2}}{2\mu(t)}
 +\frac{\hat{p}_x\hat{p}_y}{2\nu(t)}
 +\lambda(t)\hat{y}^2\right]  |\phi_{\theta}(t)\rangle,
\end{eqnarray}
with
\begin{eqnarray}
\frac{1}{\mu(t)}=\frac{1}{2u_x^{2}}+\frac{1}{2u_y^{2}}, \qquad
\frac{1}{\nu(t)}=\frac{1}{u_y^{2}}-\frac{1}{u_x^{2}}, \qquad
\lambda(t)=\frac{\eta(t)u_x  u_y}{\sqrt{2}}.
\end{eqnarray}
Note that the Hamiltonian in equation (\ref{separada}) shows that the operators involved in the variable $\hat{x}$ commute as they are simple powers of $\hat{p}_x$, such that the equation (\ref{separada}) is readily solvable as this operator act as a $c$-number for the variable $\hat{y}$. Therefore we have been able to split the Hamiltonian into two a term that is a free particle in $\hat{x}$ (time dependent) and a TDHO in $\hat{x}$ with an extra term, damping, proportional to
$\hat{p}_x$.

In order to take the equation above to a more familiar form, we  transform, with the unitary operator $\hat{D}=\exp\{i\alpha(t)\hat{p}_y\} \exp\{\beta(t)\hat{y}\}$, the above equation, namely $\ket{\phi_{D}}=\hat{D}\ket{\phi_{\theta}}$ we obtain the final and solvable form of the Hamiltonian
\begin{eqnarray}\label{last}
i\frac{\partial|\phi_{D}(t)\rangle}{\partial t}=\left[\frac{\hat{p}_y^{2}}{2\mu(t)}+\lambda(t)\hat{y}^2+\frac{\hat{p}_x^{2}}{2\mu(t)}
 -\frac{\beta\hat{p}_x}{2\nu(t)}
 +\alpha\dot{\beta}+\lambda\alpha^2+\frac{\beta^2}{2\mu}\right]  |\phi_{D}(t)\rangle,
\end{eqnarray}
where $\alpha$ and $\beta$ are functions not only of time but of the momentum operator $\hat{p}_x$ and obey the system of differential equations
\begin{eqnarray}
\dot{\alpha}+\frac{\beta}{\mu}-\frac{\hat{p}_x}{2\nu}=0, \qquad \dot{\beta}-2\lambda\alpha=0.
\end{eqnarray}
We can note that the Hamiltonian in equation (\ref{last}) has been separated in two parts: one of them a time dependent harmonic oscillator that depends only on $\hat{y}$ and $\hat{p}_y$ (and powers of them) and therefore there are Ermakov-Lewis methods to solve it and the other part that depends only on $\hat{p}_x$ (and its powers) and therefore, it is integrable \cite{JPA}.

Finally, it is worth to mention how the different transformations act on wavefunctions. It is not difficult to show that
\begin{equation}
\hat{R}_\theta \phi\left(x,y \right) =\phi\left[ \frac{\cos\left( \theta\right)x-\sin\left( \theta\right) y }{\cos^2\left( \theta\right) },
\sin\left( \theta \right) x+\cos\left( \theta \right) y  \right],
\end{equation}
where  $\phi(x,y)$ is an arbitrary, but well behaved, function of $x$ and $y$. 

To study the action of the $\hat{R}_\theta $ operator over an arbitrary function $F\left(x,y \right) $, we make
\begin{equation}
\hat{R}_\theta=\exp\left[ i\theta\left( \hat{x}\hat{p}_y - \hat{y}\hat{p}_x\right) \right] 
=\hat{T}_y \hat{S}_{xy} \hat{T}_x,
\end{equation}
where
\begin{align}
\begin{split}
\hat{T}_y=&\exp\left[ i \tan \theta  \hat{x}\hat{p}_y\right],
\\ 
\hat{S}_{xy}=&\exp\left\lbrace  -i \ln\left[\cos \theta\right]   \left(\hat{x} \hat{p}_x-\hat{y} \hat{p}_y\right)\right\rbrace 
\\
\hat{T}_x=&\exp\left[ -i\tan \theta  \hat{y}\hat{p}_x\right] .
\end{split}
\end{align}
Note that the operator $ \hat{S}_{xy}$ is a product of squeeze operators \cite{Loudon,Yuen,Caves,Vidiella,Barnett} in $x$ and $y$. We can prove that as
\begin{equation}\label{teo1}
\hat{T}_x F\left( x \right)=F\left[x-y\tan \theta   \right]  ,
\qquad
\hat{T}_y G\left( y \right)=G\left[y+x\tan \theta   \right] , 
\end{equation}
and the action of the squeeze operators 
\begin{equation}
\exp\left( i r  \hat{p}_x \hat{x} \right)x=\exp(2r)x, \qquad \exp\left( i r  \hat{p}_y \hat{y} \right)y=\exp(2r)y.
\end{equation}

\section{Conclusions}
We have shown that the quantum invariant for $N$-coupled time dependent harmonic oscillators is indeed constant for arbitrary restitutive oscillator time dependent functions as well as arbitrary time dependent coupling between them. We have translated this result to its classical version. In the case of two oscillators, we have shown how to solve the Hamiltonian for arbitrary functions of time, as formerly it had been solved only when the functions were related in specific ways \cite{Macedo2012}. We did it by using the orthogonal functions invariant introduced in reference \cite{JPA} that allowed us to split the Hamiltonian in such a way that it was left to solve a single time dependent harmonic oscillator, which is a well-known problem \cite{Lewis,Macedo2012}. 

The Ermakov-Lewis invariant for a single oscillator does not involve the time dependent parameters explicitly. This well known fact is not fulfilled when each coupled oscillator is considered separately. The quantum invariants $\hat{G}_{x}$ and $\hat{G}_{y}$ given by \eqref{eq:Q00 ux quant inv} and \eqref{eq:Q00 uy quant inv}, involve the coupling variable $\eta(t)$. However, the Ermakov-Lewis invariant of the whole system, in this case, the two coupled oscillators, i.e. $\hat{G}_{x}+\hat{G}_{y}$ no longer involves $\eta(t)$. Therefore, the invariant of the complete system is again  explicitly independent of the time varying parameters. This remark is also evinced for the $N$-coupled system. The invariant for $N$-coupled oscillators \eqref{Ninvariant}, is the Ermakov invariant of the complete system, it is again explicitly independent of any of the time varying parameters.

\end{document}